# A note on minimal matching covered graphs

V. V. Mkrtchyan

Department of Informatics and Applied Mathematics, Yerevan State University,
Yerevan,0025, Republic of Armenia, e-mail: vahanmkrtchyan2002@yahoo.com

*For Anush*

**Abstract**

A graph is called matching covered if for its every edge there is a maximum matching containing it. It is shown that minimal matching covered graphs contain a perfect matching.

Keywords: Maximum matching, Matching covered graph, Minimal matching covered graph, Perfect matching

Let $Z^+$ denote the set of nonnegative integers. We consider finite undirected graphs $G = (V(G), E(G))$ without multiple edges or loops and isolated vetices [2], where $V(G)$ and $E(G)$ are the sets of vertices and edges of $G$, respectively. For a vertex $u \in V(G)$ define the set $N_G(u)$ as follows:

$$N_G(u) \equiv \{e \in E(G) / e \text{ is incident with } u \}.$$

In a connected graph $G$ the length of the shortest $u-v$ path [2] is denoted by $\rho(u,v)$, where $u, v$ are vertices of the graph $G$. For a vertex $w \in V(G)$ and $U \subseteq V(G)$ set:

$$\rho(w, U) \equiv \min_{u \in U} \rho(w, u).$$

The set of all maximum matchings [2,4] of a graph $G$ is denoted by $M(G)$, and for $e \in E(G)$ define the set $M(e)$ as follows:

$$M(e) \equiv \{F \in M(G) / e \in F \}.$$

A vertex $u \in V(G)$ is said to be covered (missed) by a matching $F \in M(G)$ if $N_G(u) \cap F \neq \emptyset$ ( $N_G(u) \cap F = \emptyset$ ). A matching $F \in M(G)$ is called perfect if it covers every vertex $v \in V(G)$.

For a graph $G$ define the subgraph $C(G)$ as follows:

$$C(G) \equiv G \setminus \{e \in E(G) / \text{ for every } F \in M(G) \; e \notin F\}.$$

The graph $G$ is said to be matching covered if $G = C(G)$, and is said to be minimal matching covered if it satisfies the following condition, too:

$$G - e \neq C(G - e) \text{ for every } e \in E(G).$$

In this paper it is proved that every minimal matching covered graph contains a perfect matching.

The idea of the subgraph $C(G)$ of a graph $G$ is not new in graph theory. It stems from the idea of the core of a graph introduced in [1] and [3]. Roughly speaking, the core of a graph $G$ is the subgraph $C(G)$ if the cardinality of a maximum matching of $G$ equals to that of minimum point cover for $G$, and is the empty graph otherwise.

The same is true for the idea of matching covered graph. In the "bible" of matching theory [4] one can find a detailed analysis of the structure of 1-extendable graphs (connected matching covered graphs containing a perfect matching) and all necessary references of its development. In terms of [4] the main result of present paper can be reformulated in the following way: connected minimal matching covered graphs are 1-extendable.

Non defined terms and conceptions can be found in [2,4,5].

**Lemma.** If $G$ is a connected, matching covered graph, which does not contain a perfect matching, then

(1) for every edge $e = (u, v) \in E(G)$ there is a $F \in M(G)$ such that $F$ misses either $u$ or $v$;

(2) if for edges $e, e' \in E(G)$ $M(e) = M(e')$ then $e = e'$.

**Proof.** (1) For every $F \in M(G)$ consider the sets $A(F)$ and $B(F)$ defined in the following way:
$$A(F) \equiv \{w \in V(G) / F \text{ covers } w\},$$
$$B(F) \equiv \{w \in V(G) / F \text{ misses } w\}.$$

Clearly, for each $F \in M(G)$ the following holds:
$$V(G) = A(F) \cup B(F), A(F) \cap B(F) = \emptyset, A(F) \neq \emptyset, B(F) \neq \emptyset.$$

For an edge $e = (u, v) \in E(G)$ define a mapping $\mu_e : M(G) \to Z^+$ as follows:
$$\mu_e(F) \equiv \min\{\rho(u, B(F)), \rho(v, B(F))\}, \text{ where } F \in M(G).$$

Choose $F_0 \in M(G)$ satisfying the condition:
$$\mu_e(F_0) \equiv \min_{F \in M(G)} \mu_e(F).$$

Let us show that $F_0$ misses either $u$ or $v$. For the sake of contradiction assume $F_0$ to cover both $u$ and $v$. Let $w_0, (w_0, w_1), w_1, \ldots, w_{k-1}, (w_{k-1}, w_k), w_k$ be a simple path of the graph $G$ satisfying the conditions:
$$w_0 \in B(F_0), \{w_1, \ldots, w_k\} \subseteq A(F_0), \{w_{k-1}, w_k\} = \{u, v\}, k = 1 + \mu_e(F_0), k \geq 2.$$

Set $e' \equiv (w_1, w_2)$. Let us prove that $e' \notin F_0$. If $e' \in F_0$ then consider the matching $F_1 \in M(G)$ defined as follows:
$$F_1 \equiv (F_0 \setminus \{e'\}) \cup \{(w_0, w_1)\}.$$

It is clear that $\mu_e(F_1) < \mu_e(F_0)$, which contradicts to the choice of $F_0$, therefore $e' \notin F_0$. Take a maximum matching $F_0' \in M(e')$ satisfying the condition:
$$|F_0 \cap F_0'| = \max_{F' \in M(e')} |F_0 \cap F'|.$$

Let us show that $w_0 \in A(F_0')$. If $w_0 \notin A(F_0')$ then assume:
$$F_0'' \equiv (F_0' \setminus \{e'\}) \cup \{(w_0, w_1)\}.$$

Note that $F_0'' \in M(G)$ and $\mu_e(F_0'') < \mu_e(F_0)$, which is impossible, therefore $w_0 \in A(F_0')$. It is not hard to see that the choice of $F_0'$ implies that there is a simple path $v_0, (v_0, v_1), v_1, \ldots, v_{2l-1}, (v_{2l-1}, v_{2l}), v_{2l}$ $(l \geq 1)$ of the graph $G$ satisfying the conditions:

$$\{(v_0,v_1),\ldots,(v_{2l-2},v_{2l-1})\} \subseteq F_0', \{(v_1,v_2),\ldots,(v_{2l-1},v_{2l})\} \subseteq F_0,$$
$$e' \notin \{(v_0,v_1),\ldots,(v_{2l-2},v_{2l-1})\},$$
$$v_0 = w_0, v_{2l} \in \{w_1,w_2\}.$$

Set:
$$\tilde{F}_0 \equiv (F_0 \setminus \{(v_1,v_2),\ldots,(v_{2l-1},v_{2l})\}) \cup \{(v_0,v_1),\ldots,(v_{2l-2},v_{2l-1})\}.$$

Clearly $\tilde{F}_0 \in M(G)$ and $\mu_e(\tilde{F}_0) < \mu_e(F_0)$, which contradicts to the choice of $F_0$, therefore $F_0$ misses either $u$ or $v$.

(2) Suppose $e,e' \in E(G)$, $e = (u,v)$ and $e \neq e'$. Let us show that $M(e) \neq M(e')$. Take a matching $F_1 \in M(G)$ missing either $u$ or $v$. For the sake of definiteness let us assume that $F_1$ covers $u$ and misses $v$. If $e' \in F_1$ then $M(e) \neq M(e')$, therefore without loss of generality we may assume that $e' \notin F_1$. As $F_1$ covers $u$, then there is a $w \in V(G)$ such that $(u,w) \in F_1$. Set:
$$F_2 \equiv (F_1 \setminus \{(u,w)\}) \cup \{(u,v)\}.$$

Clearly, $F_2 \in M(G)$, $e \in F_2$ and $e' \notin F_2$, therefore $M(e) \neq M(e')$. The proof of the **Lemma** is complete.

From the results [1,3] and **Lemma** we have the following interesting result:

**Corollary.** Let $G$ be a connected, bipartite, matching covered graph and let $(U,W)$ be the bipartition of the set $V(G)$. If there is a $w_0 \in W$ and a $F_0 \in M(G)$ such that $F_0$ misses $w_0$, then for every $w \in W$ there is a $F \in M(G)$ such that $F$ misses $w$.

**Theorem.** Suppose that the graph $G$ satisfies the following two properties:

(1) $G$ is a matching covered graph,

(2) $G - e$ is not a matching covered graph for every edge $e \in E(G)$.

Then the graph $G$ has a perfect matching.

**Proof.** Without loss of generality we may assume $G$ to be connected. Let us show that there are two distinct edges $e$ and $e'$ such that $M(e) = M(e')$.

Take an arbitrary edge $e_0 \in E(G)$. Suppose that the edges $e_0,\ldots,e_k$ ($k \geq 0$) are already defined, and consider the graph $G - e_k$. As it is not a matching covered graph, then there exists an edge $\tilde{e} \neq e_k$ such that $M(e_k) \supseteq M(\tilde{e})$. Set $e_{k+1} \equiv \tilde{e}$.

Consider the infinite sequence $\{e_k\}_{k=0}^{\infty}$ of edges of the graph $G$. Clearly, there are numbers $i,j \in Z^+$, $i < j$ such that $e_i = e_j$. The construction of the sequence $\{e_k\}_{k=0}^{\infty}$ implies that
$$M(e_i) \supseteq M(e_{j-1}) \supseteq M(e_j) = M(e_i), \text{ and } e_{j-1} \neq e_j,$$
therefore
$$M(e_{j-1}) = M(e_j).$$

**Lemma** implies that $G$ has a perfect matching. The proof of the **Theorem** is complete.

**Acknowledgement:** I would like to thank my Supervisor R. R. Kamalian for his constant attention and support during all those years that we worked together.